\documentclass[prl,aps,showpacs,twocolumn,preprintnumbers]{revtex4}
\usepackage{amsfonts}
\usepackage{amsmath,dsfont}
\usepackage{amssymb}
\usepackage{graphicx}
\usepackage{dcolumn}
\usepackage{amsmath}
\usepackage[english]{babel}
\usepackage{amsmath,amssymb}
\usepackage{graphicx}

\begin{document}

\title{Klein tunneling in carbon nanostructures: a free particle dynamics in disguise }
\author{\textsf{V\'{\i}t Jakubsk\'y${}^1$,
Luis-Miguel Nieto${}^2$, and
Mikhail S. Plyushchay${}^{2,3,*}$}
 \\
{\small \textit{${}^1$Nuclear Physics Institute, \v Re\v z near Prague, 25068, Czech Republic}}\\
{\small \textit{${}^2$Departamento de F\'{\i}sica Te\'orica,
At\'omica y \'Optica, Universidad de Valladolid, 47071,
Valladolid, Spain}}\\
{\small \textit{${}^3$Departamento de F\'{\i}sica, Universidad de
Santiago de Chile, Casilla 307, Santiago 2, Chile}} \\
\date{}
}

\begin{abstract}
The absence of backscattering in metallic nanotubes as well
as perfect Klein tunneling in potential barriers in
graphene are the prominent electronic characteristics of
carbon nanostructures. We show that the phenomena can be
explained by a peculiar supersymmetry generated by a first
order Hamiltonian and zero-order supercharge operators.
Like the supersymmetry associated with second order
reflectionless finite-gap systems, it relates here the
low-energy behavior of the charge carriers  with the
free-particle dynamics.

\end{abstract}

\maketitle
Graphene is a genuine two dimensional  material
composed of the carbon atoms that form a honeycomb lattice.
Three of the valence
electrons of each carbon atom participate in the
interatomic interaction, while the fourth one
contributes to the conductivity of the crystal. Graphene
has been studied theoretically for a long time, see e.g.,
\cite{Wallace},
\cite{Semenoff}. However, its experimental observation
\cite{grapheneobserved} triggered a real boom of both
theoretical and experimental analysis \cite{graphene1},
\cite{graphene2}, \cite{graphene3}.

The material manifests extraordinary electronic properties,
which are the
consequence of an unusual dynamics of the low-energy charge
carriers. It was pointed out in \cite{Semenoff} that the
tight-binding description of the system is reduced to the
massless Dirac equation in the low-energy approximation. This
makes graphene an ideal test field for
$(2+1)$-dimensional QED \cite{graphene3}; due to the low Fermi
velocity $v_F$, $c/v_F\sim 300 $, it is possible to
simulate relativistic effects in condensed matter
systems which would
be unreachable experimentally otherwise.

It was predicted \cite{Klein}, \cite{SuSiuChou} that the
scattering of the relativistic electrons on the potential
barrier is qualitatively different from the nonrelativistic
case. The particles can tunnel the barrier without
reflection, provided that its height tends to infinity.
This is in contrast to the nonrelativistic regime where the
tunneling would be exponentially suppressed
\cite{SuSiuChou}. This phenomenon, known as Klein
tunneling, is not experimentally realizable with elementary
particles nowadays due to the extreme electric field needed
to observe the predicted difference between relativistic
and nonrelativistic scattering \cite{KatsnelsonKlein}.

The scattering of the low-energy quasiparticles in graphene
on the barrier with translational symmetry in one dimension
was analyzed in \cite{KatsnelsonKlein}, \cite{Falco},
\cite{Pereira}. The absence of backscattering was noticed
for normal incidence.  The effect is independent of the
height of the barrier and, hence, is testable
experimentally \cite{Kleinconfirmed}. A similar phenomenon
was observed earlier \cite{experimentnanotuba} and
discussed theoretically \cite{Todorov}, \cite{Ando1},
\cite{Mceuen} in the context of electron transport in
carbon nanotubes. The perfect transmission of the
low-energy charge carriers occurs in metallic nanotubes
despite the presence of a scattering potential generated by
impurities. The absence of backscattering was understood as
a consequence of topological singularity identified with a
Dirac point, see \cite{Ando2}, \cite{semi}, or as a result
of the pseudospin conservation \cite{KatsnelsonKlein}.

We provide here a \textit{simple, alternative}
explanation for the absence of backscattering in the
carbon nanostructures within the \textit{framework of
supersymmetric quantum mechanics}. We shall discuss
a broad class of potentials
in graphene as well as in the metallic nanotubes
with the range exceeding the interatomic distance.

The honeycomb lattice is a superposition of two
triangular sublattices, $A$ and $B$. The
eigenstate $\Psi$ of the Hamiltonian can be then written as
$\Psi=c_A \Psi_A + c_B\Psi_B$, where $\Psi_A$ and
$\Psi_B$ are atomic wave functions of the
sublattices whereas $c_A$ and $c_B$ are slowly varying
amplitudes. $\Psi$ is a Bloch function, which acquires
a nontrivial phase factor when shifted by a
translation vector $\mathbf{R}$ of the Bravais lattice,
 $\Psi (\mathbf{k},\mathbf{x}+\mathbf{R})=e^{i \mathbf{k\, R}}\Psi(\mathbf{k},\mathbf{x}).$
Fermi surface of graphene is formed by discrete points.
There are six of them in the first Brillouin zone, situated
in its corners, see Fig.\ref{fig1}. In the analysis of the
low-energy behavior of the charge carriers, it is sufficient
to consider just two of them, denoted as Dirac points
$\mathbf{K}$ and $\mathbf{K'}= -\mathbf{K}$. The remaining
four Dirac points do not represent distinct electronic states.
They can be obtained either from $\mathbf{K}$ or $\mathbf{K'}$ by
translations in the reciprocal lattice.

\newsavebox{\figpet}
    \savebox{\figpet}{
    \scalebox{1}{
    \includegraphics[scale=.6]{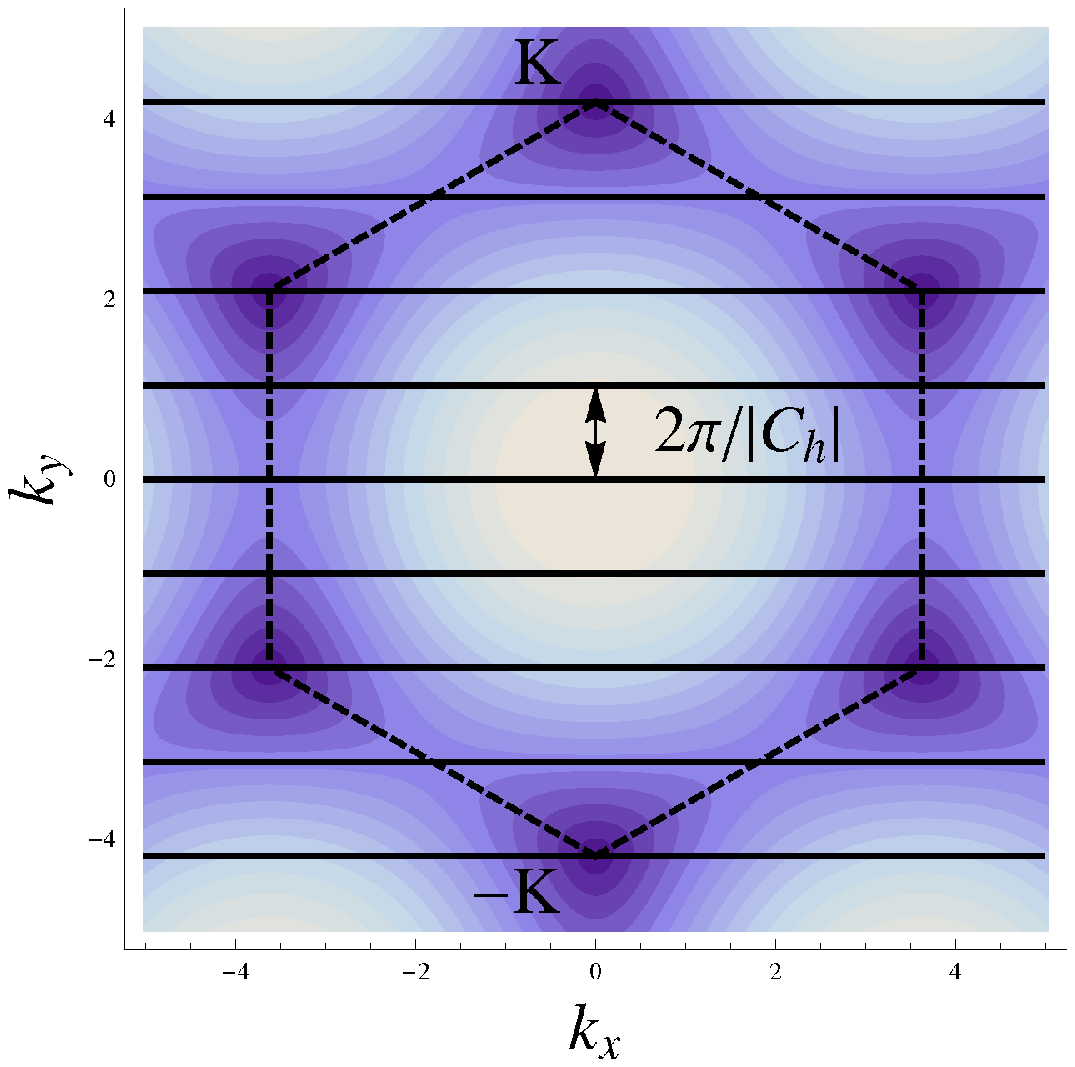}
    }
    }
\begin{figure}[h!]\begin{center}
 \usebox{\figpet}
\caption{(color online). The hexagon represents the first
Brillouin zone of graphene. The blue contours show constant
values of the energy modulus $|E|$, $E=\pm\gamma_0\sqrt{3+2\cos\mathbf{k
a_1}+2\cos\mathbf{k a_2}+2\cos\mathbf{k (a_1-a_2)}}$,
$\gamma_0=2(\sqrt{3}a)^{-1}v_F$. The primitive translation
vectors of the hexagonal lattice are fixed as $\mathbf{a_1} = a(0, 1)$
and $\mathbf{a_2} = a/2\left(\sqrt{3},1\right)$,
while $\mathbf{K} = a^{-1}(0,{4\pi}/{3}$),
$a=\sqrt{3}d$, $d$ is the nearest bond length of the graphene lattice
(we put here $a=1$ and $v_F=\sqrt{3}/2$).
The deep blue corresponds to the lowest values of $|E|$.
The energy surface forms deep valleys near the Dirac points.
The parallel lines illustrate the allowed momenta for the
zig-zag nanotube $(6,0)$, i.e. $\mathbf{k\, C_h}/2\pi\in\mathbb{Z}$,
\cite{CBR}.
}\label{fig1}
\end{center}\end{figure}

In the vicinity of Dirac points, the behavior of the
system is described by the massless Dirac equation. When
the effective Hamiltonian is
considered in the valley of the point $\mathbf{K}$
and expanded up to the terms linear in the momenta,
the energy eigenvalue equation acquires the form
\cite{Semenoff} (we put $\hbar=1$)
\begin{equation}\label{KDirac}
H\psi=-i\,v_F\, ({\sigma_1\partial_x+\sigma_2\partial_y})
\,\psi=E\psi,
\end{equation}
where $\sigma_{1,2}$ are Pauli matrices. Spinor
$\psi$ reads explicitly
$\psi=(c_A,c_B)^te^{i\mathbf{\delta
k x}},$ where $\mathbf{\delta k=k-K}$ and $t$ is a transposition.
Degree of freedom associated with the components $c_A$ and $c_B$
is called pseudospin, see \cite{graphene1}, \cite{graphene2}.

A single wall carbon nanotube can be created by rolling up
and gluing appropriately a single  graphene strip \cite{CBR}.
The circumference (chiral) vector $\mathbf{C_h}$ is an integer multiple
of the primitive translation vectors  $\mathbf{a_1}$
and $\mathbf{a_2}$ of the Bravais lattice, $\mathbf{C_h}=n_1\mathbf{a}_1+n_2\mathbf{a}_2$. It
defines uniquely the nanotube \cite{foot1} by the
periodic condition imposed on the wave functions,
$\Psi(\mathbf{x}+\mathbf{C_h})=\Psi(\mathbf{x})$.
Let us suppose that the coordinate system is chosen such
that the $y$ axis points in
the direction of the chiral vector,
$\mathbf{k}\,\mathbf{C_h}=k_y|\mathbf{C_h}|$. Taking into
account Bloch character of the wave functions, one can
see that $\Psi$ has the required periodicity as long as
$\mathbf{k}\,\mathbf{C_h}=k_y|\mathbf{C_h}|= 2\pi q$ for an integer $q$.
Hence, momentum $\mathbf{k}$ is quantized in the
$y$-direction and the allowed values
form equidistant lines in the first Brillouin zone, with a
step $\frac{2\pi}{|{\bf C_h}|}$, see Fig.\ref{fig1}\,.

There are two qualitatively different positions of the
Dirac point $\mathbf{K}$ with respect to the allowed lines,
in coherence with two main classes of carbon nanotubes.  In
case of metallic nanotubes, $\mathbf{K}$ is crossed by one
of the lines. For semiconducting nanotubes there is a
minimum distance $min|\mathbf{k}-\mathbf{K}|
=2\pi/|3\mathbf{C_h}|$ between the Dirac point and the
closest line  \cite{nanotrubkyvlastnosti}, which leads to
the opening of the gap between the valence and conduction
bands.

The low-energy behavior of the charge carriers in
the nanotube with the chiral vector $\mathbf{C_h}$
is approximated by
\begin{equation}\label{Hnano}
H_{\epsilon}\psi= v_F\left(-i \sigma_1\partial_x+ \epsilon \sigma_2\right)\psi=E\psi,
\end{equation}
where the value of $\epsilon\in\left\{0,\pm
\frac{2\pi}{3|\mathbf{C_h}|}\right\}$ depends on the type
of the nanotube\,: it is zero for metallic nanotubes and
nonvanishing for semiconducting ones. The parameter
$\epsilon$ can be alternatively regarded as a fictitious
magnetic flux \cite{KaneMele} or the mass of the
quasiparticle. Thus, the $H_0$, given by (\ref{Hnano}) with
$\epsilon=0$, coincides with the Dirac Hamiltonian of the
free massless particle in one dimension.

The real-world nanostructures are not perfect. The crystal
can have impurities; or it can be in presence of an
external field. We suppose that these effects are
represented by a potential $V$, which is vanishing at
infinity, varies smoothly on the scale of the interatomic
distance and is of the range larger than it (and, hence,
does not cause the intervalley transition of the electronic
states). Then it can be consistently incorporated into
(\ref{KDirac}) and (\ref{Hnano}) as the unit matrix
multiplied by $V$ \cite{Ando1}. The stationary evolution
equation for metallic nanotubes acquires a form
\begin{equation}\label{impurity}
 H_{V}\psi=(-iv_F \sigma_1\partial_x+V(x))\psi=E\psi.
\end{equation}
The same Hamiltonian $H_V$ can be obtained as a reduction
of the Hamiltonian of graphene with a potential barrier,
$H+V$, as long as potential is translationally invariant in
the $y$-direction, i.e. $V=V(x)$,
\begin{equation}\label{Hv}
H_{V}=e^{-i\delta k_y y}(H+V)e^{i\delta k_y y}|_{\delta k_y=0}.
\end{equation}
In this case, $H_V$
represents the energy operator of the particle
with normal incidence ($\delta k_y=0$) on the barrier.
The Hamiltonian $H_V$ describes reflectionless system,
independently on the particular form of the potential $V$.
This peculiar fact was explained by the presence
of Berry phase in the wave functions, which causes
the Born series for
backscattering to vanish identically \cite{Ando2}.

\vskip0.2cm Let us look, however, at the problem from a new
perspective. In nonrelativistic quantum mechanics, the
reflectionless systems play an important role in the theory
of solitons and are intimately related to the nonlinear
integrable systems \cite{Crum}. Their particular properties
come hand in hand with the existence of the supersymmetry,
that is based on the Darboux-Crum transformations
\cite{Crum}, \cite{Khare}, \cite{PoschlTeller}. The
supercharges intertwine such systems with the free-particle
model and stay behind the absence of backscattering in the
conduction band.

This rises the question whether
the perfect tunneling in the
carbon nanostructures has a
similar algebraic background.
The answer is affirmative, despite the fact that the
relativistic Hamiltonian $H_V$ is of the first order,
contrary to the second order Hamiltonians of the
nonrelativistic reflectionless systems
\cite{foot2}.

Let us define the hermitian operators
\begin{equation}\label{matop1}
 \mathcal{H}=\left(\begin{array}{cc}H_{V}&0\\0&H_{0}\end{array}\right),\quad
 \Gamma=\left(\begin{array}{cc}\mathbf{1}&0\\0&-\mathbf{1}\end{array}\right),
\end{equation}
where $\mathbf{1}$ is the unit two-by-two matrix.
Besides the operator $\Gamma$, the Hamiltonian
$\mathcal{H}$ has two other symmetries,
\begin{equation}\label{matop2}
 \mathcal{U}_1=\left(\begin{array}{cc}0&U^{\dagger}\\U&0\end{array}\right),\quad
\mathcal{U}_2=i\,\Gamma\,\mathcal{U}_1,
\end{equation}
where $U=U(x)$ is a unitary operator of a local chiral
rotation (chiral gauge transformation),
\begin{equation}\label{u}
 U=e^{i\alpha\sigma_1}=\cos\alpha\, \mathbf{1} +
 i\sin\alpha\,\sigma_1,\quad U^{\dagger}=U^{-1},
\end{equation}
dependent on the interaction potential,
\begin{equation}
 \alpha(x)=\frac{1}{v_F}\int^x V(\tau)d\tau.
\end{equation}
They satisfy relations
\begin{equation}\label{susyrel}
 [\mathcal{H},\mathcal{U}_a]=0,\quad
\{\mathcal{U}_a,\mathcal{U}_b\}=2\delta_{ab}\mathds{1},\quad a,b=1,2.
\end{equation}
The anticommutator of $\mathcal{U}_1$ and $\mathcal{U}_2$
is a zero-order polynomial in $\mathcal{H}$,
to be proportional to the central
element $\mathds{1}$  which is the unit
$4\times 4$ matrix. The
grading operator $\Gamma$ classifies the
operators (\ref{matop1}) and $\mathds{1}$ as bosonic
(they commute with $\Gamma$) while operators (\ref{matop2}) are
fermionic (they anticommute with $\Gamma$).

Relations (\ref{susyrel}) constitute the $N=2$ zero-order
supersymmetry extended by the central charge $\mathds{1}$
and graded by $\Gamma$. Like in the nonrelativistic case of
a reflectionless system with the $n$-gap, second order
Hamiltonian (where the order $n$ of supersymmetry is fixed
by the number $n$ of bound states \cite{PoschlTeller}),
this structure underlies the absence of the backward
scattering in the system given by the first order
Hamiltonian $H_V$. The relation $UH_{V}=H_{0}U$, implied by
the commutator in (\ref{susyrel}) and the unitarity of $U$,
reveal the unitary equivalence of $H_V$ with the free
massless Dirac Hamiltonian $H_0$. This proves the absence
of backscattering\,: the setting given by $H_V$ is unitary
equivalent to the free massless particle system and, hence,
it shares its trivial scattering properties.

Because of this peculiar supersymmetric structure, based on
the nontrivial unitary equivalence, all the integrals of
the free-particle system represented by $H_0$ have their
analogs in the system given by $H_V$. In particular, the
pseudospin is conserved in $H_V$ just because of
$[H_0,\sigma_1]=[U,\sigma_1]=0$. The free-particle momentum
$-i\partial_x$ transforms into the integral $-i\partial_x
+\frac{1}{v_F}\sigma_1 V(x)\equiv -i\mathcal{D}_x$ for
$H_V$, that has a form of a covariant derivative with a
chirality-dependent charge. Then $H_V$, like $H_0$, is
presented as a composition of the integrals, $H_V=-iv_F
\sigma_1\mathcal{D}_x$.

The unitary equivalence of $H_V$ with the free-particle
system is broken when the nonvanishing effective mass (the
coefficient of $\sigma_2$) is present in the impurity
Hamiltonian. It emerges in semiconducting nanotubes ($H_{V}
= H_\epsilon +V$ with $m=\epsilon =\pm 2\pi/|3{\bf C_h}|$),
or when other than normal incidence of the particles is
considered in graphene system (\ref{Hv}), $H_V =e^{-i\delta
k_y y}(H + V )e^{i\delta k_y y}|_{\delta k_y=m\neq 0}$
$=H_m+V $, where $H_m=H_0+v_Fm\sigma_2$. In these cases,
the unitary transformation of $H_V$ yields $UH_V U^{-1}
=H_0 +mv_F \sigma_2(x)$, $\sigma_2(x)= U\sigma_2U^{-1}=\cos
2\alpha\,\sigma_2-\sin 2\alpha\,\sigma_3$, and
\begin{equation}\label{susybreaking}
 UH_V=H_mU - 2v_F m \sin \alpha\, \sigma_3.
\end{equation}
Therefore, the scale of supersymmetry breaking in the
massive case is of the order of $m$, and the contribution
of the potential is controlled by the factor $|\sin
\alpha|\leq 1$. Hence, for the close-to-the-normal
incidence ($m=\delta k_y\sim 0$), the potential barrier
remains almost perfectly transparent for any $V(x)$. This
is coherent with \cite{Falco}, where the scattering on
$n-p$ junction was analyzed. For general values of the
effective mass, the scattering properties of $H_V$ are
nontrivial, however, and depend on the explicit form of the
potential. The quasiparticles can be confined in graphene,
the charge carriers get localized in semiconducting
nanotubes, see \cite{Mceuen}, \cite{confined}.

Zero-order supersymmetry, based on a unitary equi\-valence
of the superpartner Hamiltonians, can be formally
constructed for any quantum system. It is sufficient to
make a unitary transformation of an initial Hamiltonian to
get its superpartner. However, the supersymmetry is
nontrivial and manifests its predictive power when the
Hamiltonians describe different physics, like in
(\ref{matop1}). There, it provided a simple explanation for
the absence of backscattering in the considered carbon
nanostructures. As an example of another application, it
may be used to determine the $s$-wave of the Dirac operator
in polar coordinates \cite{graphene3},
$H(r,\theta)=H_0(r)-iv_F\sigma_2 \frac{1}{r}\partial_\theta
+V(r)$, where $H_0(r)=iv_F\sigma_1(\partial_r
+\frac{1}{2r})$, for any potential $V(r)$. Indeed,
$H(r,\theta)$ acts on the subspace of $s$-waves
($\psi=\psi(r)$) as $H_0(r)+V(r)$. We can use an
$r$-dependent unitary mapping to get rid of the interaction
term and to obtain the operator $H_0(r)$ with analytically
computable eigenstates.

In the context of graphene, the supersymmetry appeared
earlier in the analysis of the zero-energy states in
presence of external magnetic field \cite{Novikov}, and in
the study of the quantum Hall effect in particular
\cite{Ezawa}. In that case, the \textit{first order}
supercharges are proportional to the Dirac Hamiltonian,
while the supersymmetric Hamiltonian is of the
\textit{second order}. The structure in (\ref{susyrel}) is
completely different: the supersymmetric Hamiltonian
$\mathcal{H}$ is of the \textit{first order} and the
existence of the \textit{zero-order} supercharges
$\mathcal{U}_a$ provides a complete information on the
eigenstates of $H_V$. Together with (\ref{susybreaking}),
it suggests to analyze the system with small effective mass
perturbatively.

The algebraic framework presented here can be broadened in
different ways. For instance, the apparent similarity with
the nonrelativistic reflectionless systems suggests to
extend the analysis by employing the Darboux-Crum
transformations in the context of the higher-order
(nonlinear) supersymmetry \cite{footnote2,foot2}. We
believe that in such a generalized form the supersymmetry
can serve as a useful tool in the study of the low-energy
excitations of the charge carriers in the wrapped graphene,
where the dynamics is governed by the Dirac Hamiltonian in
a curved space. Further discussion on this problem, and
other possible applications indicated above goes beyond the
scope of the present article.

\vspace{3mm}
\noindent
 The work of MSP has been partially supported by
 FONDECYT Grant 1095027, Chile and  by Spanish Ministerio de
 Educaci\'on under Project
SAB2009-0181. LMN has been partially supported by the
Spanish Ministerio de Ciencia e Innovaci\'on under Project
MTM2009-10751 and Junta de Castilla y Le\'on (Excellence
Project GR224). VJ was supported by the Czech Ministry of
Education, Youth and Sports under Project LC06002 and by
GA\v CR Grant P203/11/P038. MSP and VJ thank the Department
of Physics of the University of Valladolid for hospitality.

\vspace{3mm}
${}^*mikhail.plyushchay@usach.cl$

\end{document}